\documentclass{aa}  

\usepackage{graphicx}
\usepackage{txfonts}
\begin{document}

   \title{SHALOS: Statistical \textit{Herschel}-ATLAS Lensed Objects Selection}
   \titlerunning{SHALOS}


   \author{J. Gonz\'{a}lez-Nuevo \inst{1}, S.L. Su\'{a}rez G\'{o}mez\inst{2}, L. Bonavera\inst{1},
            F. S\'{a}nchez-Lasheras\inst{2}, F. Argueso\inst{2}, L. Toffolatti\inst{1}, \\D. Herranz\inst{3},
            C. Gonz\'{a}lez-Guti\'{e}rrez\inst{4}, F. Garc\'{i}a Riesgo\inst{5}
          \and
          F. J. de Cos Juez\inst{5}
          }
    \authorrunning{Gonz\'{a}lez-Nuevo et al.}

   \institute{Departamento de F\'{i}sica, Universidad de Oviedo, C. Federico Garc\'{i}a Lorca 18, 33007 Oviedo, Spain\\
              \email{gnuevo@uniovi.es}
         \and
         Departamento de Matem\'{a}ticas, Universidad de Oviedo, C. Federico Garc\'{i}a Lorca 18, 33007 Oviedo, Spain
         \and
         Instituto de Física de Cantabria (CSIC-Universidad de Cantabria), Avda. de los Castros s/n, E-39005 Santander, Spain
         \and
         Departamento de Inform\'{a}tica, Universidad de Oviedo, C. Federico Garc\'{i}a Lorca 18, 33007 Oviedo, Spain
         \and
         Departamento de Explotaci\'{o}n y Prospecci\'{o}n de Minas, Universidad de Oviedo, Oviedo, 33004 Asturias, Spain
             }


 
  \abstract
   {The statistical analysis of large sample of strong lensing events can be a powerful tool to extract astrophysical and/or cosmological valuable information. However, the number of such events is still relatively low, mostly because of the lengthily observational validation process on individual events.}
   {In this work we propose a new methodology with a statistical selection approach in order to increase by a factor of $\sim 5$ the number of such events. Although the methodology can be applied to address several selection problems, it has particular benefits in the case of the identification of strongly lensed galaxies: objectivity, minimal initial constrains in the main parameter space, preservation of the statistical properties.}
   {The proposed methodology is based on the Bhattacharyya distance as a measure of the similarity between probability distributions of properties of two different cross-matched galaxies. The particular implementation for the aim of this work is called SHALOS and it combines the information of four different properties of the pair of galaxies: angular separation, luminosity percentile, redshift and optical/sub-mm flux density ratio.}
   {The SHALOS method provided a ranked list of strongly lensed galaxies. The number of candidates for the final associated probability, $P_{tot}>0.7$, is 447 with an estimated mean amplification factor of 3.12 for an halo with a typical cluster mass. Additional statistical properties of the SHALOS candidates, as the correlation function or the source number counts, are in agreement with previous results indicating the statistical lensing nature of the selected sample.}
   {}


   \maketitle
%

\section{Introduction}
\label{sec:intro}
In the last decade, surveys at sub-millimetre (sub-mm) wavelengths have revolutionized our understanding of the formation and evolution of galaxies by revealing an unexpected population of high-redshift, dust-obscured galaxies called sub-mm galaxies (SMGs) which are forming stars at a tremendous rate \citep[i.e. star formation rate, SFR$\gtrsim1000~M_\odot yr^{-1}$; ][]{Bla99}. Data collected before the advent of the European \textit{Herschel} Space Observatory \citep[\textit{Herschel}; ][]{Pil10} and the South Pole Telescope \citep{Cal11}, suggested that the number density of SMGs drops off abruptly at relatively bright sub-mm flux densities ($\sim50$ mJy at $500~\mu m$), indicating a steep luminosity function and a strong cosmic evolution for this class of sources.

Several authors have argued that the bright tail of the sub-mm number counts may contain a significant fraction of strongly-lensed galaxies \citep[SLGs;][]{Bla96,Neg07}. The \textit{Herschel} Multi-tiered Extragalactic Survey \citep[HerMES; ][]{Oli12} and the \textit{Herschel} Astrophysical Terahertz Large Area Survey \citep[H-ATLAS;][]{Eal10} are wide-field surveys ($\sim\!380$ deg$^2$ and $\sim\!610$ deg$^2$, respectively) conducted by the \textit{Herschel} satellite. Thanks to their sensitivity and frequency coverage both surveys have led to the discovery of several lensed SMGs (\citet{Neg17} and references therein). The selection of SLGs at these wavelengths is made possible by the steep number counts of SMGs \citep{Bla96,Neg07}; in fact, almost only those galaxies whose flux density has been boosted by an event of lensing can be observed above a certain threshold, namely $\sim\!100$ mJy at $500~\mu m$. Similarly, at mm wavelengths, the SPT survey has already discovered several tens of SLGs \citep[e.g.][]{Vie13,Spi16} and other lensing events have been found in the Planck all-sky surveys \citep{Can15,Har16}.

With \textit{Herschel} data, \citet{Neg10} produced the first sample of five SLGs by means of a simple selection in flux density at $500~\mu m$. Preliminary source catalogues derived from the full H-ATLAS were then used to identify the sub-mm brightest candidate lensed galaxies for follow-up observations with both ground based and space telescopes to measure their redshifts (see \citet{Neg17} with 80 SLG candidates and references therein) and confirm their nature \citep{Neg10,Bus12,Bus13,Fu12,Cal14}. Using the same methodology, i.e. a cut in flux density at $500~\mu m$, \citet{War13} have identified 11 SLGs over 95 deg$^2$ of HerMES, while, more recently, \citet{Nay16} have published a catalogue of 77 galaxies candidate at lenses with $S_{500~\mu m}\gtrsim100$ mJy extracted from the HerMES Large Mode Survey \citep[HeLMS][]{Oli12} and the \textit{Herschel} Stripe 82 Survey \citep[HerS; ][]{Vie14}, over an area of 372 deg$^2$. Altogether, the extragalactic surveys carried out with \textit{Herschel} are expected to deliver a sample of $\sim\!200$ of sub-mm bright SLGs. 

Moreover, as argued by \citet{GN12}, this number might increase to over a thousand if the selection is based on the steepness of the luminosity function of SMGs \citep{Lap12} rather than that of the number counts. The HALOS (\textit{Herschel}–ATLAS Lensed Objects Selection) method relies on the fact that SLGs tend to dominate the brightest end of the high-$z$ luminosity function. This method was demonstrated by looking for close associations (within 3.5 arcsec) with VIKING galaxies \citep{Fle12} that may qualify as being the lenses after a primary selection based on \textit{Herschel} photometry ($S_{350\mu m}>85$mJy, $S_{250\mu m}>35$mJy, $S_{350\mu m}/S_{250\mu m}>0.6$ and $S_{500\mu m}/S_{350\mu m}>0.6$). To be conservative, the candidates were further restricted to objects whose VIKING counterparts have redshifts z>0.2. After comparing both SLGs candidate lists, it was shown that about 70\% of SMGs with luminosities in the top 2\% percentile were also identified with the second method.

Although the HALOS method is a step forward to increase the number of SLGs candidates, its conclusions are based on a sample with very restrictive selection criteria that makes difficult to extrapolate its performance to a more general case. Moreover, the main parameter of the method, the top luminosity percentile, does not have a clear optimal value and the choice of such value makes the method very subjective.

For the above reasons and taking into account the slow pace of follow-up campaign confirmation, in this work we propose a new methodology based on the similarity between probability distributions of pairs of galaxies from two different catalogues (one for the potential foreground galaxies acting as lenses and another one for the potential background sources), associated to a set of observables as the redshift or the angular separation. The characteristics of this method make it more objective and easily reproducible, providing a final probability ranked list of SLGs candidates. Moreover, with very few initial constraints, the statistical properties of the SLGs candidates are not biased and can be studied statistically before the observational confirmation of each individual case. The data sets and the initial selection criteria are presented in Section \ref{sec:cats}. The general methodology is discussed in Section \ref{sec:methodology}, while the details for the particular implementation of the general methodology to the identification of SLGs and the main results are described in Section \ref{sec:shalos}. Some of the statistical properties of the SHALOS SLGs candidates are estimated and discussed in Section \ref{sec:induction}. Finally the main conclusions are presented in Section \ref{sec:conclusions}.

\section{Data}
\label{sec:cats}
In this work we use the official H-ATLAS catalogues, the largest area extragalactic survey carried out by the \textit{Herschel} space observatory \citep{Pil10}. With its two instruments PACS \citep[Photoconductor Array Camera and Spectrometer;][]{Pog10} and SPIRE \citep[Spectral and Photometric Imaging Receiver;][]{Gri10} operating between 100 and 500 $\mu$m, it covers about 610 deg$^{2}$. The survey is comprised of five different fields, three of which are located on the celestial equator \citep[GAMA fields or G09, G12 and G15;][]{Val16,Bou16,Rig11,Pas11,Iba10} covering in total an area of 161.6 deg$^2$. The other two fields are centred on the North and South Galactic Poles \citep[NGP and SGP fields;][]{Smi17,Mad18, Fur18} covering areas of 180.1 deg$^2$ and 317.6 deg$^2$, respectively. As described in detail in \citet{Bou16}, for the GAMA fields, and \citet{Fur18}, for the NGP field, a likelihood ratio method was used to identify counterparts in the Sloan Digital Sky Survey \citep[SDSS; ][]{Aba09} within a search radius of 10 arcsec of the H-ATLAS sources with a $4\sigma$ detection at $250~\mu m$. We were not able to use the SGP field in this work because there is no overlap with the SDSS survey.

We are going to focus on those sources with a cross-matched optical counterpart and, therefore, there is an implicit $4\sigma$ detection at $250~\mu m$ initial selection criteria. In addition, we discard sources flagged as stars and those galaxies without an optical redshift estimation.

\subsection{Sub-mm photometric redshifts}
\label{sec:photoz}
Photometric redshift are provided in the H-ATLAS catalogues but they are based on the optical cross-matched information. This means that if the cross-matched sources are different galaxies, the estimated redshifts tend to correspond to the ones at lower redshift. We have used the spectroscopic redshift when available.

To have an independent estimation of the redshift for the potential SMGs (the high redshift counterparts) we follow the usual approach to derive the sub-mm photometric redshifts. Following previous works \citep{Lap11,Pea13,GN12,GN14,Ivi16,GN17,Bon19}, the sub-mm photometric redshifts were estimated by means of a minimum $\chi^2$ fit of a template SED to the SPIRE data (using PACS data when possible). The SED of SMM J2135-0102 \citep[`The Cosmic Eyelash' at $z = 2.3$;][]{Ivi10,Swi10} is known to be the best overall template to describe the SMGs population, at least for $z>0.8$. When comparing with spectroscopic redshifts, the usage of this template provides the best performance with a minimum difference dispersion: $\Delta z/(1 + z) = -0.07$ and a dispersion of 0.153 \citep{Ivi16,GN12,Lap11}.

In order to obtain more reliable sub-mm photometric redshifts, we restrain ourselves to those sources with at least $3\sigma$ photometric estimations at 350 and $500~\mu m$. Moreover, we further focus only on those sources with estimated sub-mm photometric redshifts with $z>0.8$.

Finally, using the estimated photometric redhsift and the SED of SMM J2135-0102 we calculate the Bolometric Luminosity for each of the SMGs.

\section{Methodology}
\label{sec:methodology}
For our purpose we need a method to compare two different probability distributions deriving a quantity that can be interpreted as a probability in order to combine the information obtained from different comparisons. Among the different defined statistical distances between distributions, we find that the Bhattacharyya distance fulfill our requirements.
In statistics, the Bhattacharyya distance \citep{Bat43} measures the similarity of two discrete or continuous probability distributions. It is closely related to the Bhattacharyya coefficient  \citep[BC;][]{Bat43}, i.e. the overlap estimate of two probability distributions. Among other applications, the Bhattacharyya distance is widely used in research of feature extraction and selection \citep[e.g,][]{Ray89, Cho03}. 

The Bhattacharyya distance for two continuous probability distributions $p$ and $q$ can be expressed as:
\begin{equation}
  D_B (p,q) =-ln(BC(p,q))=-ln\left(\int dx\sqrt{p(x)q(x)}\right),  
\end{equation}
\noindent
where BC(p,q) denotes the Bhattacharyya kernel or Bhattacharyya Coefficient, 
with $0\leq D_B \leq \infty$ and $0\leq BC\leq1$. When $p$ and $q$ are two normal distributions, the Bhattacharyya distance can be computed as:
\begin{equation}
\label{eq:2normal}
    D_B(p,q) = \frac{1}{4} ln\left( \frac{1}{4} \left( \frac{\sigma^2_p}{\sigma^2_q} + \frac{\sigma^2_q}{\sigma^2_p} + 2\right)\right) + \frac{1}{4} \left( \frac{(\mu_p-\mu_q)^2}{\sigma_p^2+\sigma_q^2}\right),
\end{equation}
\noindent
where $\sigma^2_p$ ($\sigma^2_q$) is the variance of the $p$ ($q$) distribution and $\mu_p$ ($\mu_q$) is the mean of the $p$ ($q$) distribution.

The usage of the Bhattacharyya distance, or distance in general, is a novel approach in the identification of specific source characteristics or events based on cross-matched pair of galaxies. Moreover, it has some advantages with respect to more traditional approaches:
\begin{itemize}
\item The calculation of a distance between two probability distributions relies just on the parameters describing such distributions and are determined by observations (such as beam size, positional uncertainty, redshift uncetainties) and, therefore, do not require any assumption on previous knowledge, \textit{priors}, or limits.
\item As an extension of the previous point, the usage of the distance avoids complicated calculations of several probabilities of the samples and model parameters, most of which depend on \textit{a priori} assumptions, required in Likelihood Ratio (LR) approaches \citep[e.g., ][]{Bou12}. LR was successfully implemented to cross-match optical catalogues where the assumptions and probability or parameter estimations were reasonably acceptable. However, the implementation of the LR technique to cross-match catalogues observed in different wavelengths bands becomes very complex.
\item There are other statistical methods for similarity measurements between probability distributions. Most of them consist on statistical hypothesis tests, as for example the the two sample t-test. However, they work with $p$-values, a measure of the statistical evidence for the validation of certain hypothesis, which is usually misunderstood and wrongly used as a measurement of probability. On the contrary, the BC gives a similarity measurement that can be safely interpreted as a probability.
\item Moreover, if we have two or more similarity measurements it is not clear how to combine the $p$-values obtained from each measurement. In the distance approach, the combination of different similarity measurements is straightforward, being just the multiplication of the $BC$ estimated values (similar to the general rule in probabilities).
\item Finally, it should be commented that also the Bayesian alternative to classical hypothesis testing has some limitations. The usage of the Bayes factors can be individually applied to each observational property, but it rises the issue on how to combine the "strength of the evidence" for each individual observable.
In general, Bayes factors are used as a Bayesian model comparison methodology (a generalization of the LR technique). With this approach an ideal model has to be defined to be compared with and it requires knowledge about prior distributions \citep{Bud08,Bud11}. For our purpose, such characteristics make the Bayesian alternative a limited or biased approach.
Some improvements were introduced to overcome these limitations as the Intrinsic Bayes Factor presented in \citet{Ber96}, but it requires the estimation of intrinsic priors and over-complicates the calculation.
\end{itemize}

Therefore, we propose the combination of various distance measurements between two probability distributions (associated to different observable quantities related to the pair of galaxies) as a new simple, objective (without any prior and based on observational probability distributions), modular (additional information can be added as any time to review the overall final probabilities) and flexible (can be adapted for different purposes: identifying strong lensing events, discriminate sub-populations, star-galaxy classification, etc) methodology to identify particular kind of sources or events by cross-matching different catalogues. A natural extension of this methodology could be to implement a neural network to be trained to perform the same task, as already used in other contexts \citep[e.g., ][]{Ode92,Sto92}.

\section{SHALOS}
\label{sec:shalos}

\begin{figure}[tbp]
\centering 
\includegraphics[width=\columnwidth]{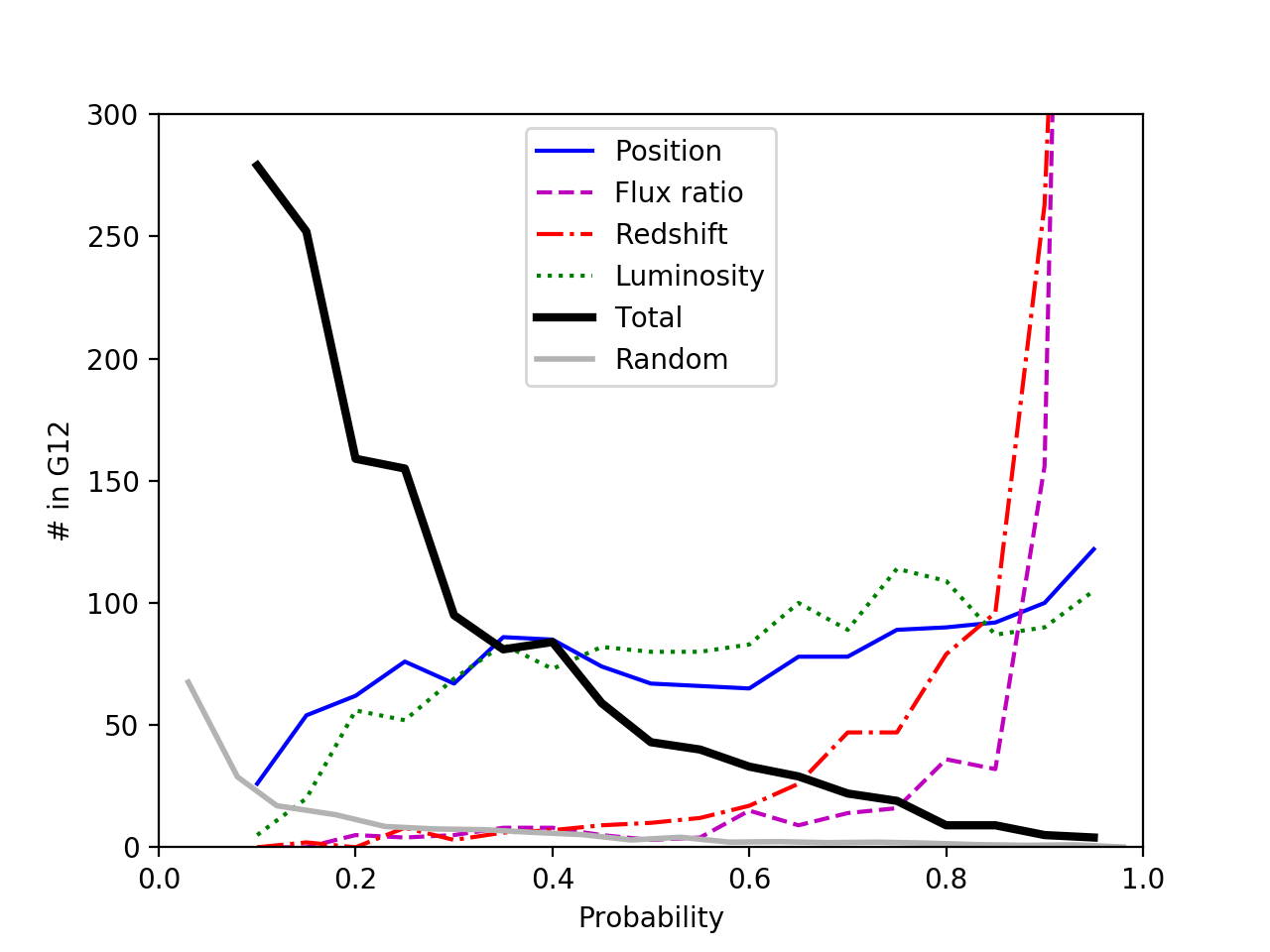}
\caption{\label{fig:P_hist} Comparison of the variation of the number of sources with the different probabilities associated to the observables considered in this work. The total probability is shown as a thick black line while the estimated probability of random pairs is shown as a grey line.}
\end{figure}

Our main objective in this work is to identify a list of the most probable cases of a strong lensing event between optical galaxies acting as lenses and SMGs acting as sources. We named as SHALOS\footnote{`shalo' in the modern urban English slang means "share your location", that it is also adequate to our purpose.} ("Statistical \textit{Herschel}-ATLAS Lensed Object Selection") the specific implementation details to this particular scientific task of the methodology in sec. \ref{sec:methodology}.
The intention of SHALOS is to produce a probability ranked list of potential SLGs. This ranked list try to be as much objective as possible and easily reproducible.  

In general, gravitational lensing events between two different samples share the following characteristics: two different closed objects (small angular separation) at different distances (different redshifts) with the background flux density amplified with respect the rest of the source population (higher luminosity). Therefore, we focus on the following observables: angular separation, optical vs sub-mm flux density comparison, redshift and  background luminosity.

\begin{itemize}
    \item \textit{Angular separation ($BC_{pos}$)}.- The closer in the sky is the lens-source pair, the higher is the gravitational lensing probability. For this observable we compare the positional uncertainty distributions of each pair of galaxies described as gaussian distributions centred in the galaxy position with a dispersion equal to the positional uncertainty (eq. \ref{eq:2normal}). In this case, a higher overlap, i.e. higher BC value, implies a potential higher lensing probability. The global astrometric RMS precision of SDSS is $\sim0.1$ arcsec\footnote{https://www.sdss.org/dr12/scope/} while it is $\sim 2.4$ arcsec for H-ATLAS catalogues \citep{Bou16,Fur18}. Due to the huge difference between the probability distribution dispersion for both cases, the maximum overlap is $BC_{pos}\sim0.3$, for a zero angular separation. Therefore, for aesthetic purposes (i.e., in order to have the best candidates near $BC_{pos}\sim1.0$), we normalize the $BC_{pos}$ to the maximum overlap value.
    \item \textit{Redshift ($1-BC_z$)}.- In this case, we are interested in objects at different redshifts and, therefore, with a minimum redshift probability distribution overlap, i.e. $1-BC_z$. Similar to the previous case, we compare the redshift uncertainties of a pair of galaxies described as gaussian distributions centred in the lens/source redshfit best values with a dispersion equal to the redshift uncertainties.
    As the source has to be at higher redshift than the lens, any residual of the source redshift probability distribution at lower redshift than the lens one is considered also part of the overlap. In particular,  we consider as overlap any residual probability distribution area of the source galaxy that were at lower redshift than $<\mu_{z,lens} - 3\sigma_{z,lens}$, being $\mu_{z,lens}$ and $\sigma_{z,lens}$ the mean readshift and its associated gaussian dispersion. This modification became important when dealing with spectroscopic redshift with very small redshift uncertainties. For the lens candidates, the uncertainty is 0.01 when a spectroscopic redshift measurement is available or the $1\sigma$ error for the photometric estimations. For the sources, the uncertainty is the maximun value between the photometric estimation method $1\sigma$ error or the statistical one, 0.153 \citep{Ivi16}.
    \item \textit{Optical vs. sub-mm flux density ratio ($1-BC_r$)}.- To help to distinguish if the source/lens galaxies are the same one, we also consider the ratio between the optical $r$ band and the sub-mm $350~\mu m$ flux densities. This additional information can be useful when the redshifts are similar (typically $z\sim 0.8-1.0$). For each matched galaxy pair, we estimate the flux densities ratio and its uncertainty and we compared it with the expected one from \citet{Smi12}, a stacked SED for typical galaxies at $z<0.5$. If the measured ratio is similar to the stacked SED, the matched galaxies are probably the same galaxy with redshift $z<0.8$. Therefore, as in the redshift case, we are interested in those cases with minimum probability distributions overlap, i.e. $1-BC_r$.
    \item \textit{Luminosity percentile ($L_{perc}$)}.- A source galaxy amplified due to a strong lensing effect will tend to have higher Luminosity with respect to other galaxies at similar redshift \citep{GN12}. The bolometric luminosity of each source galaxy canditate is compared with those at similar redshift ($\mu_z-\sigma_z < z < \mu_z+\sigma_z$): the higher the associated percentile the more probable is the hypothesis of a strong lensing event. Taking into account the results from \citet{GN14, GN17} and \citet{Bon19}, most of the event candidates will be produced by weak lensing with typical amplifications below 50\%. In these cases, we expect luminosity percentiles fluctuating around $\sim0.5$.
\end{itemize}

Finally, we combine the information from the four observables to obtain a total strong lensing probability associated to each SLGs candidate:
\begin{equation}
    P_{tot} = BC_{pos} * (1 - BC_z) * (1 - BC_r) * L_{perc}
\end{equation}

\subsection{SHALOS produced catalogues and usage}

The SHALOS methodology can be applied to any pair of catalogues and start the cross-matching process from scratch. However, we decided to apply it using the cross-match information already in the official H-ATLAS catalogues (see sec. \ref{sec:cats}). 

There are some pros and cons to this decision. On the one hand, the H-ATLAS cross-match was limited to pairs of objects within angular distance $<10$ arcsec. In addition, when multiple counterparts were possible the LR technique was used to chose the most probable one. We consider that initializing the SHALOS methodology using a pair list limited to angular separation $<10$ arcsec does not introduce any bias in identifying SLGs: taking into account the typical positional uncertainties, separation distances larger than this limit are severely penalized within the proposed methodology. This is not true anymore when trying to study the weak lensing regime, with potential gravitational effects at even larger angular separation, depending on the lens mass.
On the other hand, the H-ATLAS catalogues provide not only spectroscopic redshift (when available), but also photometric ones for most of the optical counterparts, that we would have had to compute otherwise.

Overall, using the H-ATLAS catalogues provided us the opportunity to compare our results with the LR ones. This comparison is very interesting allowing a discussion on the differences between both methodologies and their optimal applicability cases.

Therefore, for each entry in one of the H-ATLAS catalogues with a cross-matched optical galaxy, we estimate the associated $P_{tot}$ as described before (Sec. \ref{sec:shalos}). Then all the entries with $P_{tot}<0.1$ are removed and the remaining ones are sorted by their $P_{tot}$ associated value in decreasing order. The SHALOS catalogues can be found as the online material of the publication. From the official H-ATLAS catalogues, we have maintained the most critical information: name, the \textit{Herschel} flux densities and r magnitude, angular separation, LR Reliability and the optical spectroscopic and photometric redshift. Then we added the SHALOS intermediate information as sub-mm redshifts, bolometric luminosity, the four observables associated probabilities and the final total one.

We foresee the usage of the produced SHALOS catalogues mainly as the ranked input list of sub-mm strong lensing targets for follow ups campaigns with high resolution facilities as the HST, Keck or ALMA. The SHALOS ranked list can be used to easily select the `best' event candidates that complies with the observational campaign criteria as sky region, flux density limits, redshift range, etc.

However, the SHALOS method is an approach based mainly on observable measured quantities with minimal assumptions and minimal \textit{a priori} limits. 
This means that we can statistically consider most of the top ranked selected events as real and safely perform their analysis, also comparing with previous results on this field. This comparison can be used as a validation by induction of the SHALOS method and new results can be obtained with respect to previous analysis (that are based only on confirmed events).

\subsection{SHALOS results}

\begin{table*}
\caption{PCA loadings for each of the considered observables ($BC_{pos}$, $(1-BC_r)$, $(1-BC_z)$, $L_{perc}$), and their correspondent influence on each of the components.}             
\label{table:PCA1}      
\centering                          
\begin{tabular}{c | c c c c c}        
\hline\hline                 
  & Comp.1 & Comp.2 & Comp.3 & Comp.4 \\    
 \hline      
$(1-BC_r)$ & -0.07567495 & -0.004521857 & -0.050656304 & -0.99583472 \\
$BC_{pos}$  &  0.81154329 & -0.560949927 & -0.155269678 & -0.05122494 \\
$(1-BC_z)$   &  -0.12034156 & 0.093595030 & -0.986553689 & 0.05890413 \\
$L_{perc}$ & -0.56673512 & -0.822529454 & -0.006089687 &  0.04711173 \\
\hline                                   
\end{tabular}
\end{table*}

\begin{table*}
\caption{PCA components relevancy considering standard deviation, proportion of explained variance and the cumulative proportion of variance.}             
\label{table:PCA2}      
\centering                          
\begin{tabular}{c | c c c c c}        
\hline\hline                 
  & Comp.1 & Comp.2 & Comp.3 & Comp.4 \\    
 \hline      
Standard deviation     & 0.2646 & 0.2086 & 0.1212 & 0.09205 \\
Proportion of Variance & 0.5122 & 0.3183 & 0.1075 & 0.06198 \\
Cumulative Proportion  & 0.5122 & 0.8305 & 0.9380 & 1.00000 \\
\hline                                   
\end{tabular}
\end{table*}

Most of the detected galaxies in the H-ATLAS catalogue do not even have an optical counterpart within $10$ arcsec. As a consequence they are not considered by the SHALOS method. From this point, we will focus our work only on those event candidates with a $P_{tot}> 0.1$ at least. We consider that below such value of the associated probability there is a completely negligible probability for the event of being a SLG.

Figure \ref{fig:P_hist} summarizes the behaviour of the four probabilities, related to the observable quantities previously described, considered in the SHALOS method. The variation of the number of selected galaxies with respect to the associated probability is an indication of their relative importance. The redshift (dot-dashed red line) and flux ratio (dashed magenta line) observables are introduced to assure that the pair of galaxies are different objects at different distances. They are not very restrictive because the criteria used to select the initial sample was already able to discard potential dubious pairs and low redshift candidates. Their effect is more important for those cases with background redshift near the imposed lower limit, $z>0.8$.

On the contrary, the luminosity percentile (dotted green line) and the angular separation (blue line) information are the most restrictive. As anticipated by the HALOS method \citep{GN12}, the first one will help to select those candidates with higher probability of a stronger gravitational lensing effect. The positional one simply prefer the closest pairs, that normally translate into higher lensing amplifications.

The total associated probability, $P_{tot}$, is shown as a thick black line indicating that the number of SLGs decreases with $P_{tot}$, as expected. The estimated number of potential random pairs that fulfill all the methodology criteria is shown as a grey solid line: for $P_{tot}>0.1$ it can be considered negligible. It was estimated by maintaining the same lens galaxy sample and simulating the background sources. The simulated background sample mimic the real background sample statistics \citep[redshift distribution; source number counts at $250~\mu m$, \citet{Lap11}; `The Cosmic Eyelash' SED,][]{Ivi10} but with random positions. Then, we applied the same selection sample criteria and we cross-matched it with the lens sample using the same 10 arcsec as the maximum angular distance radius. Finally, for each of the random event candidates we applied the SHALOS methodology to obtain the associated total probability, shown in Fig. \ref{fig:P_hist}. This process was repeated 10 times to derive a mean value for each $P_{tot}$ and its dispersion.

Similar conclusions can be obtained from a Principal Component Analisys (PCA). It is performed in order to set the relative relevance of the four considered probabilities by determining their separate influence on the principal components. For the PCA analysis only the $P_{tot}>0.1$ cases were considered. In the PCA, a linear combination of the (standardized) components is made to predict a certain variable: the loadings are the coefficients of this linear combination and, for each component, the sum of their squared values are the eigenvalues (components' variances).

The PCA results show that, for the two most relevant components (components 1 and 2), $BC_{pos}$ and $L_{perc}$ are the most influencing observables. In particular, $BC_{pos}$ is the most important for component 1 and $L_{perc}$ for component 2. The other two principal components corresponds almost entirely to $(1-BC_z)$ (components 3) and to $(1-BC_r)$ (component 4), whose weights are the highest in absolute value (see Table \ref{table:PCA1}).

The importance of each observable can be inferred from the proportion of variance, explained by each principal component considering the information obtained from the loadings. According to the proportion of variance shown in Table \ref{table:PCA2}, component 1 is the one that explains most of the variance (51.22\%), followed by component 2 (31.83\%). Thus, the most relevant observables are $P_{pos}$ and $L_{perc}$. The proportion of variance for component 3 and 4 is lower, and consequently the observables $(1-BC_z)$ and, mostly $(1-BC_r)$, are less important. 

\begin{figure}[tbp]
\centering 
\includegraphics[width=\columnwidth]{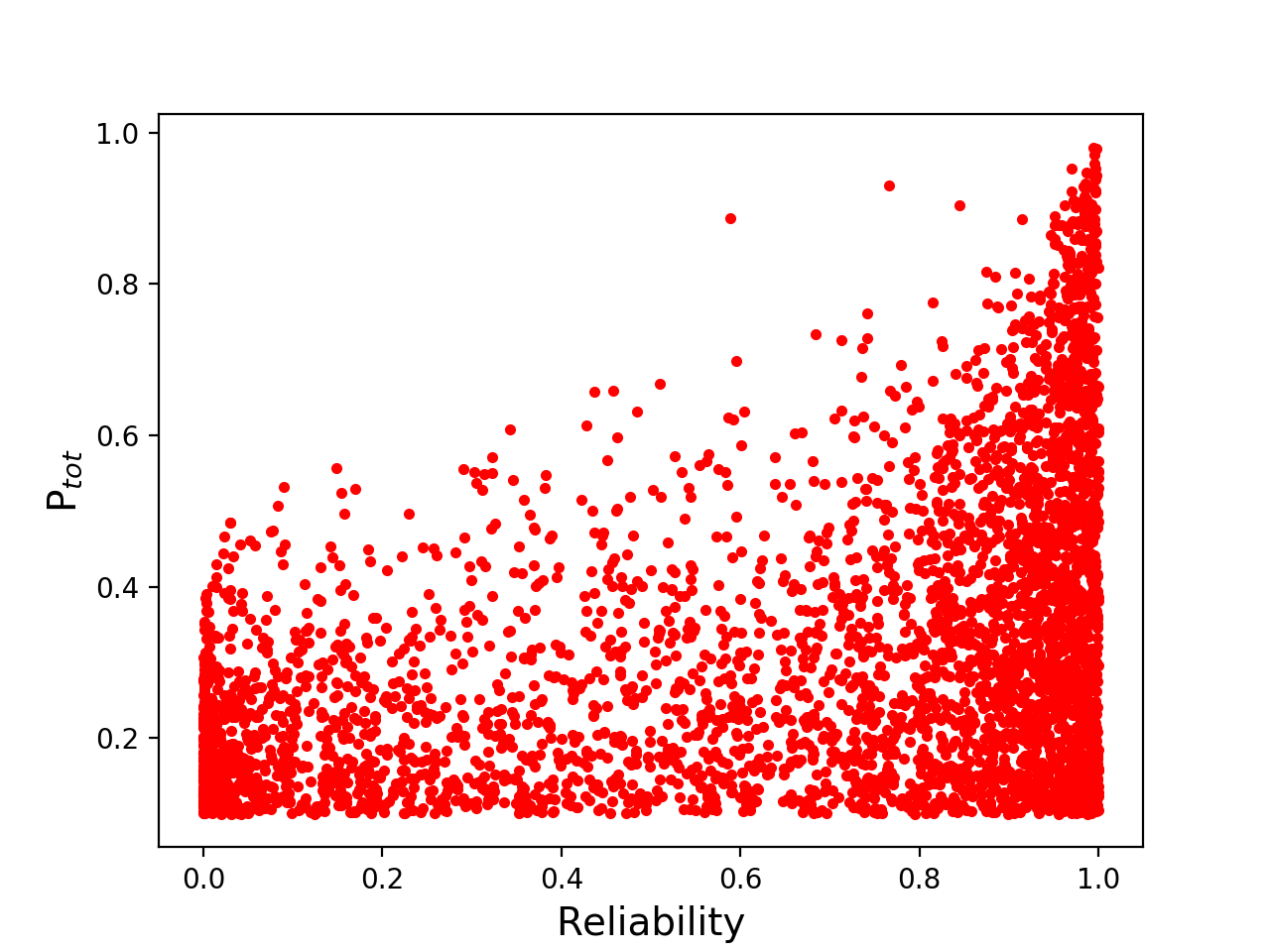}
\caption{\label{fig:P_vs_R} Comparison between P$_{tot}$ and the likelihood Reliability.}
\end{figure}

\begin{figure}[tbp]
\centering 
\includegraphics[width=\columnwidth]{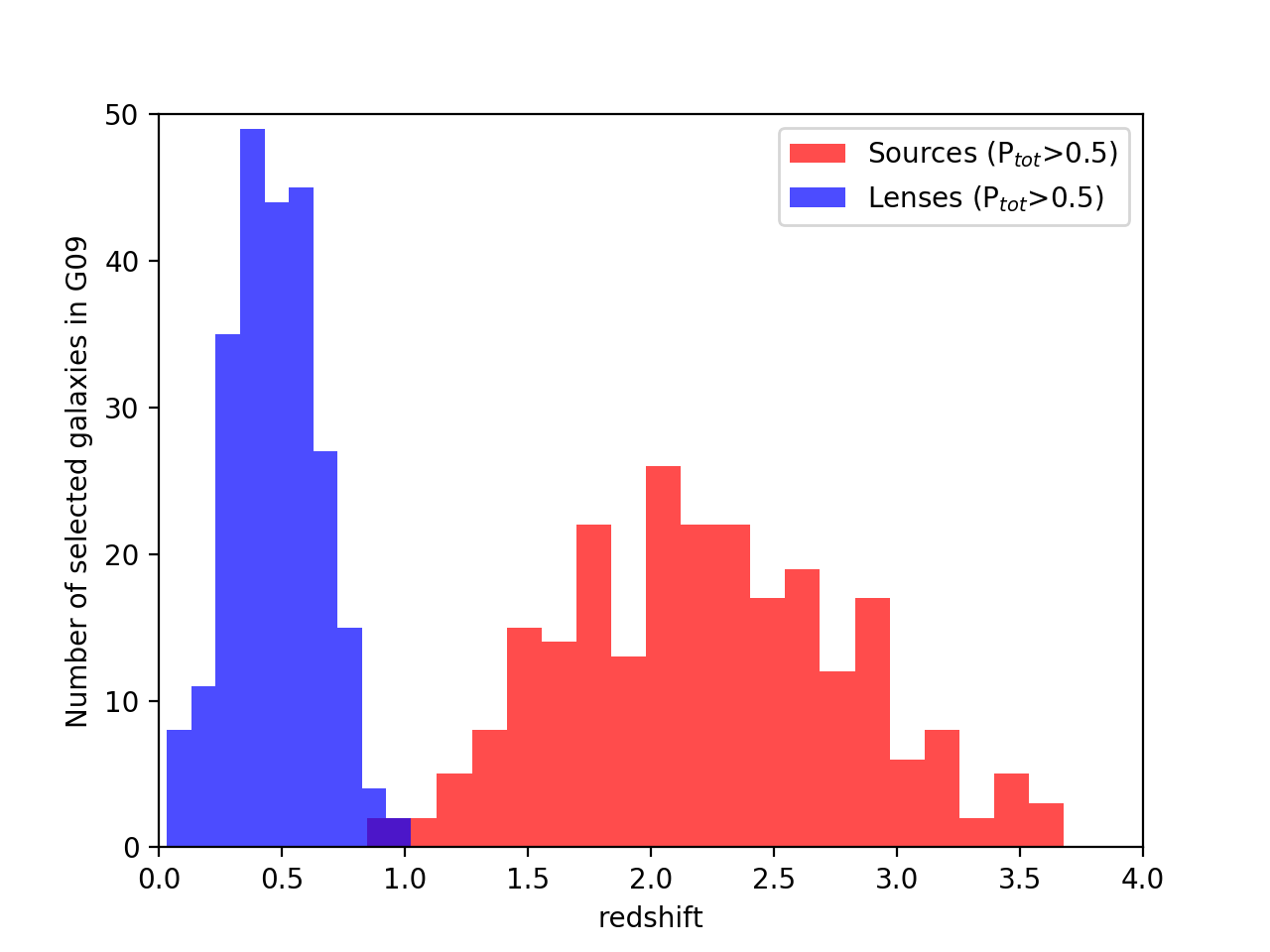}
\caption{\label{fig:dNdz} Comparison between the redshift distributions of the lenses (\textit{blue}) and sources (\textit{red}) selected by SHALOS with P$_{tot}>0.5$.}
\end{figure}

On the other hand, in Figure \ref{fig:P_vs_R} the estimated total probability for sources with P$_{tot}>0.1$ is compared with the Reliability (quantity to identify the goodness of the cross-matched SDSS local galaxies) estimated in the official H-ATLAS catalogues based on the LR cross-match approach. It is clear that both quantities differ and show almost a bimodal distribution. Cases with low Reliability values, $R<0.3$, has also relatively low associated $P_{tot}<0.5$ values. This is mainly due to the effect of the angular separation in both methodologies. However, more than half of the sources shown in Figure \ref{fig:P_vs_R} have high Reliability ($R>0.8$) with P$_{tot}> 0.1$. The reason is that those "matches" have a smaller angular separation. In the LR methodology, small angular separation results into a higher Reliability. However, at the same time, this is also one of the required characteristic in a gravitational lensing event. Without additional information, as redshift or luminosity, the LR method lacks the proper information in the case of SLGs to associate a low Reliability, as already pointed out in previous works \citep{Neg10,GN12,GN14,Bou14}.
The redshift distribution of sources (\textit{red}) and lenses (\textit{blue}) identified with P$_{tot}>0.5$ in the G09 zone are shown in Figure \ref{fig:dNdz}. The other areas have almost identical redshift distributions. The redshift distribution of the sources covers a wide range of redshift: from $\sim 0.9$ to $\sim 3.6$ with a mean value of $z\sim 2.3$. Sources below $z \simeq 1.5$ are penalized mainly due to their photometric redshift uncertainties. On the other hand, lenses show a redshift distribution with mean value of $z\sim0.5$ as expected from theoretical estimations (see \citet{Lap12} for more details) for sources around $z\sim 2.5$. It is interesting that the SHALOS method identify also several events with lenses at $z < 0.2$, because there is no initial constraint on this respect (contrary to previous works).

\begin{table*}
\caption{Summary of the SHALOS results stats.}             
\label{table:stats}      
\centering                          
\begin{tabular}{c c c c c c}        
\hline\hline                 
Zone & Initial & Sample & $P_{tot}>0.1$ & $P_{tot}>0.5$ & $P_{tot}>0.7$ \\    
 & (\#) & (\#)[\%] & (\#)[\%] & (\#)[\%] & (\#)[\%] \\
\hline      
G09 & 39660 & 2808 [7.08\%] & 1374 [3.46\%] & 240 [0.61\%] & 73 [0.18\%] \\
G12 & 38961 & 2924 [7.50\%] & 1377 [3.53\%] & 213 [0.55\%] & 68 [0.17\%] \\
G15 & 41609 & 3059 [7.35\%] & 1506 [3.62\%] & 243 [0.58\%] & 70 [0.17\%] \\
NGP & 118980 & 8437 [7.09\%] & 4129 [3.47\%] & 755 [0.63\%] & 236 [0.20\%] \\
ALL & 239210 & 17228 [7.20\%] & 8386 [3.51\%] & 1451 [0.61\%] & 447 [0.19\%] \\
\hline                                   
\end{tabular}
\end{table*}

In Table \ref{table:stats} there is a summary of the number of galaxies initially in the H-ATLAS catalogues for each of the four zones considered. It is also shown the number of selected sources at high redshift with reliable flux density measurements and the number of identified SLGs at different $P_{tot}$ values. There are already several interesting conclusions that can be extracted from these results:
\begin{enumerate}
    \item Only $\sim 7\%$ of the global H-ATLAS sources are considered reliable high redshift sources, $z>0.8$ with our current selection criteria.
    \item The results are homogeneous among the different zones, with minimal percentage variations.
    \item More than half of the high redshift selected sources have a close low-z optical counterpart and, therefore, they have a non negligible associated probability, $P_{tot}>0.1$, of being a SLG. This result is in agreement with the strong magnification bias signal measured by \citet{GN14, GN17} that implies that many of the H-ATLAS high-z sources are slightly enhanced by weak gravitational lensing.
    \item The probability of a stronger gravitational effect is boosted by increasing the $P_{tot}$ limit due to the luminosity percentile observable effect. The number of candidates  with $P_{tot}>0.5$ is greater than 1000, confirming the HALOS predictions \citep{GN12} that with more complex selection procedures it is possible to reach such numbers.
    \item Finally, the most probable candidates, $P_{tot}>0.7$, correspond to 447 (or 0.19\%) that it is $\sim5$ times the number of H-ATLAS candidates found with flux density above 100 mJy at $500\mu m$ \citep{Neg17}.
\end{enumerate}

As a check of our results, we compare SHALOS SLG candidates with those found in \citet{Neg17}. 
In the common NGP and GAMA fields, \citet{Neg17} found 50 SLG candidates, but only 32 are identified in SHALOS. The other 18 objects are excluded by us either because they have no estimated optical redshift (needed by SHALOS) or because they are flagged as star (and we are not interested in such objects).

On the one hand, following the \citet{Neg17} selection criteria, we select those SHALOS SLG candidates with a flux density at 500 $\mu m$ greater than 90 mJy and redshift greater than 0.1, to avoid very local objects. We use a lower flux density limit with respect to \citet{Neg17}, $S_{500~\mu m}\geq100$ mJy, to take into account small flux density variation in the different versions of the H-ATLAS catalogues. By applying such redshift and flux density selection in SHALOS, we obtain a list of 50 objects with $P_{tot}>0.1$, again with the common 32 SLG candidates. From the new 18 objects that are in SHALOS and not in \citet{Neg17}, 6 have $P_{tot}<0.5$, i.e. a very low probability of being actual SLGs, 8 have $S_{500~\mu m}<100$ mJy and therefore excluded from \citet{Neg17} list, 3 are identified as blazars (see Table 1 in \citet{Neg17}) and one is a local extended source (NGC5705) so that its SPIRE photometry is no reliable for the possible, if there is any, background source.

Therefore, not only is the SHALOS method as effective as the \citet{Neg17} approach for $S_{500~\mu m}>100$ mJy, but it is also able to extend the identification methodology at lower flux density limits.

\section{Validation by induction}
\label{sec:induction}

\begin{figure}[tbp]
\centering 
\includegraphics[width=\columnwidth]{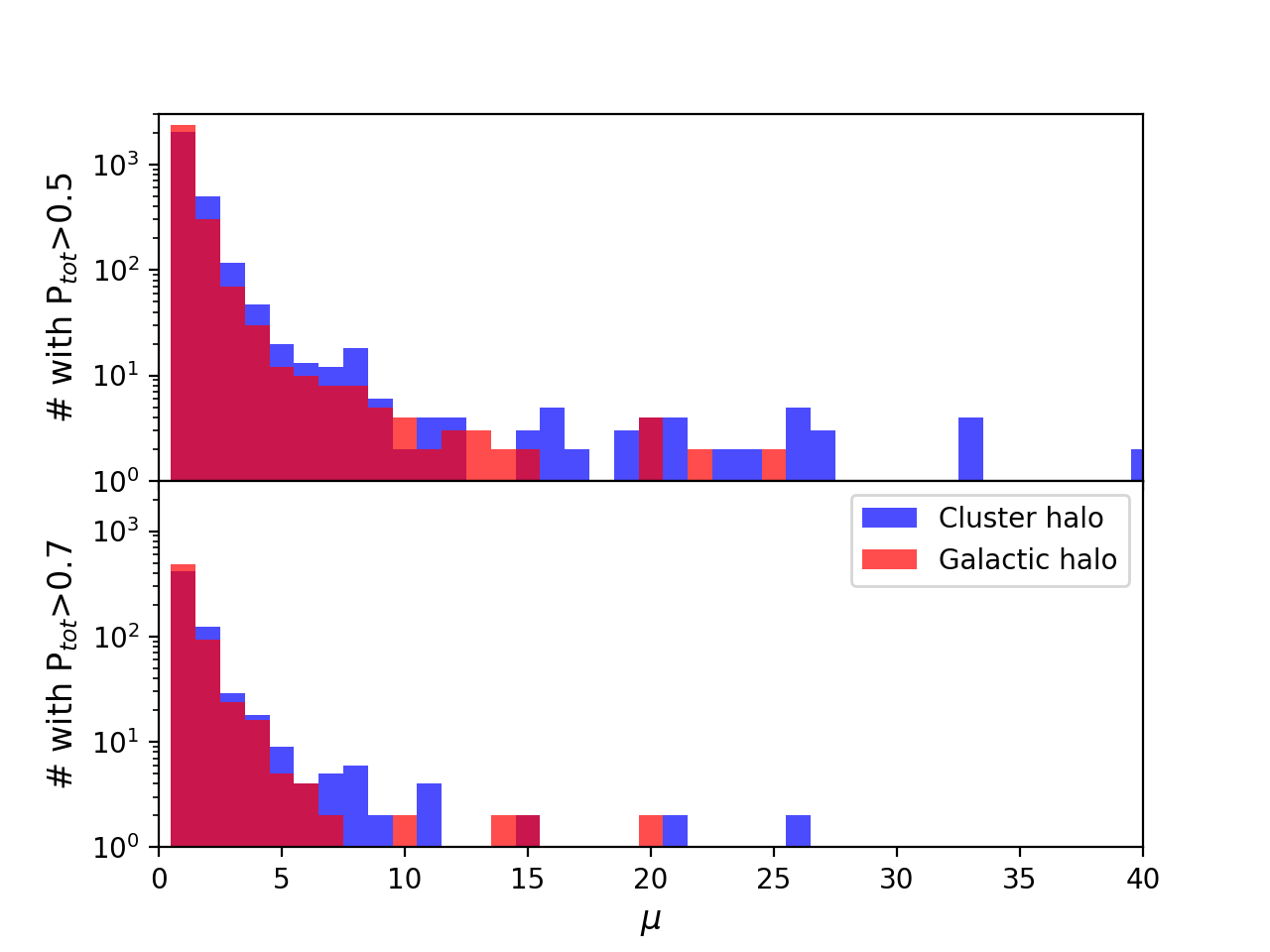}
\caption{\label{fig:mus} Tentative amplification factors derived assuming a galactic and cluster halo masses (see text for more details).}
\end{figure}

Only a follow up campaign using top instruments with high resolution and sensitivity could establish the overall performance of the proposed methodology by studying one by one each individual SLG candidate. However, even obtaining observational time in such facilities, it will take months, if not years, to built a data base large enough to derive some meaningful statistics.

For this reason, we propose an alternative and complementary approach to validate the SHALOS methodology: validation by induction. In this Section we are going to assume that all the lensing event candidates are confirmed SLGs and to study some of their statistical properties. Then, we can compare such properties with previous results or theoretical expectations to check if they are in agreement. If this is the case, we can conclude that the SHALOS provided list is mainly composed by SLGs. Therefore, we can use the SHALOS list to obtain additional valuable statistical information about this kind of events thanks to its less restrictive limits.

\subsection{Amplification factors}
The first statistical property calculated is a tentative amplification factor, $\mu$, produced by the gravitational lensing effect: there is enough information in the SHALOS list to derive an approximate $\mu$ for each of the event canditates. Following mainly the same procedure as in \citet{GN14}, we estimate for each lens the stellar mass, $M_\star$, from the r-band Luminosity, $L_r$. We considered two different scenarios: i) the gravitational lensing effect is produced mainly by the galactic halo surrounding the lens galaxy; ii) the lens galaxies are, typically, the central galaxy of a group or cluster of galaxies as indicated by the conclusions obtained by \citet{GN14} and \citet{GN17}. In this case we thus estimate a group or cluster of galaxies halo mass.

In the first case, we consider a `Singular Isothermal Sphere' (SIS) mass density profile and we can derive the galactic halo mass, $M_h$, directly from the r-band Luminosity \citep{Sha06,Ber03}:
\begin{equation}
M_h=3\times 10^{11}\left(\left(\frac{L_r}{1.3\times 10^{10}} \right)^{0.35}+ \left(\frac{L_r}{1.3\times 10^{10}} \right)^{1.65}\right)\times 10^{-0.19z}
\end{equation}

For the second scenario, we considered a `Navarro-Frank-White'(NFW) mass density profile \citep{NFW96}. The stellar mass is calculated using a modified version of the luminosity-stellar mass relationship \citep{Ber03,Ber10}:
\begin{equation}
    M_\star/L_r=3\times(L_r/10^{10.31})^{0.15} \times 10^{-0.19z},
\end{equation}
with $M_\star$ and $L_r$ in $M_\odot$ and $L_\odot$, respectively. Then the cluster halo mass is estimated by applying the stellar to halo mass relationship derived by \citet{Mos10}.

Finally, the amplification factors for both scenarios are estimated following the traditional gravitational lensing framework \citep[see for example][]{Sch05}, taking into account the derived halo masses and the source and lens redshifts \citep[we have used the concentration formula derived by][]{Pra12}.

The results, for all the different areas together, are shown in Fig. \ref{fig:mus} for two different $P_{tot}$ cuts. Even with these tentative estimations about the amplification factors, these results are encouraging. The mean (median) values of the SHALOS list for $P_{tot}>0.5$ is 1.90 (1.26) for the galactic halo case and 2.51 (1.39) for the cluster case. For a more conservative cut, with $P_{tot}>0.7$, the obtained amplification factors are on average bigger: 2.28 (1.33) for the galactic halo and 3.12 (1.47) for the cluster case. With these estimated amplification factors, the number of SHALOS candidates with $P_{tot}>0.5$ and $\mu>2$ are 17.2\% and 25.8\% for the galactic and cluster scenarios, respectively. These percentages increases to 24.3\% and 31.5\%, respectively, for the $P_{tot}>0.7$ cut.

\subsection{Auto-correlation function}
\label{sec:acorr}

\begin{figure}[tbp]
\centering 
\includegraphics[width=\columnwidth]{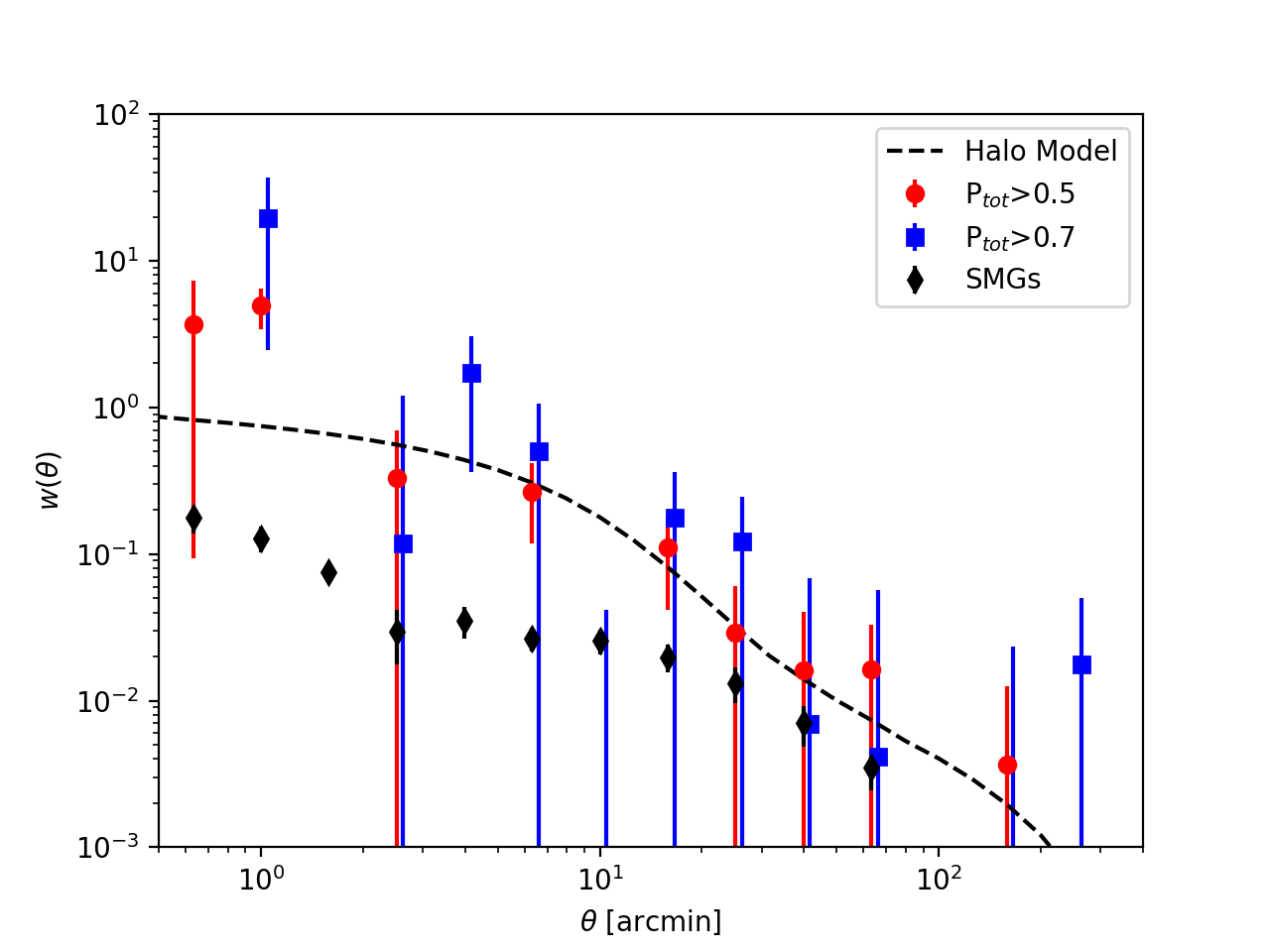}
\caption{\label{fig:ac} Auto-correlation of SHALOS SLGs with P$_{tot}>0.5 ~\& ~0.7$ compared with the theoretical estimation using the \citet{GN17} observed cross-correlation parameters. The \citet{GN17} measured auto-correlation of the H-ATLAS high-z sources (\textit{black circles}) is also shown as a comparison.}
\end{figure}

The number of SHALOS candidates is large enough to measure their two-point correlation function. If the SHALOS candidates were simply random associations their correlation function would have being negligible or noise dominated. At maximum they could have resembled the correlation of the SMGs or background sample. On the contrary, if they are real SLGs their correlation function will be in agreement with the one expected from a foreground sample with the lens derived masses.

In order to check these possibilities, we estimate the two points correlation function of the SHALOS candidates using the \citet{Lan93} estimator:
\begin{equation}
    w(\theta)=\frac{DD(\theta)- 2DR(\theta)+ RR(\theta)}{RR(\theta)},
\end{equation}
where DD, DR and RR are the normalized unique pairs of galaxies, the data-random pairs and the random-random pairs, respectively.

The measured correlation functions for $P_{tot}>$0.5 \& 0.7 are shown in Fig. \ref{fig:ac}. Although the uncertainties are significant, the SHALOS candidates have a non-zero or noise dominated correlation function. This result immediately discard the random association hypothesis. Moreover, the SHALOS candidates correlation is stronger than the measured by \citet{GN17} for the \textit{Herschel} SMGs at $z>1.2$ (\textit{black diamonds}). These galaxies are the same galaxies that constitute our background sample. Therefore, it is confirmed that there is something special about the SHALOS selected background galaxies.

Finally, we calculae, for comparison, the correlation function expected for a sample of lenses with the observed redshift distribution (blue histogram in Fig. \ref{fig:dNdz}) and the mass and halo ocupation distribution properties derived by \citet{GN17}: minimum halo mass of $\sim1.3\times 10^{13} M_\odot$, a pivotal mass to have at least une satellite galaxy of $\sim3.7\times 10^{14} M_\odot$ and the slope of the number of satellites, $\sim2$. By using the same Halo model formalism of \citet{GN17}, mainly based on \citet{Coo02}, we derived the dashed black line, in good agreement with our measured correlation functions.

Therefore, we can conclude that the angular correlation properties of the SHALOS selected candidates closely resemble the expected ones for the sample of foreground lenses. It is not a direct validation of the gravitational lensing nature of the SHALOS candidates but it is an additional statistical property that agrees with the expectations.

\subsection{Source number counts}
\begin{figure}[tbp]
\centering 
\includegraphics[width=\columnwidth]{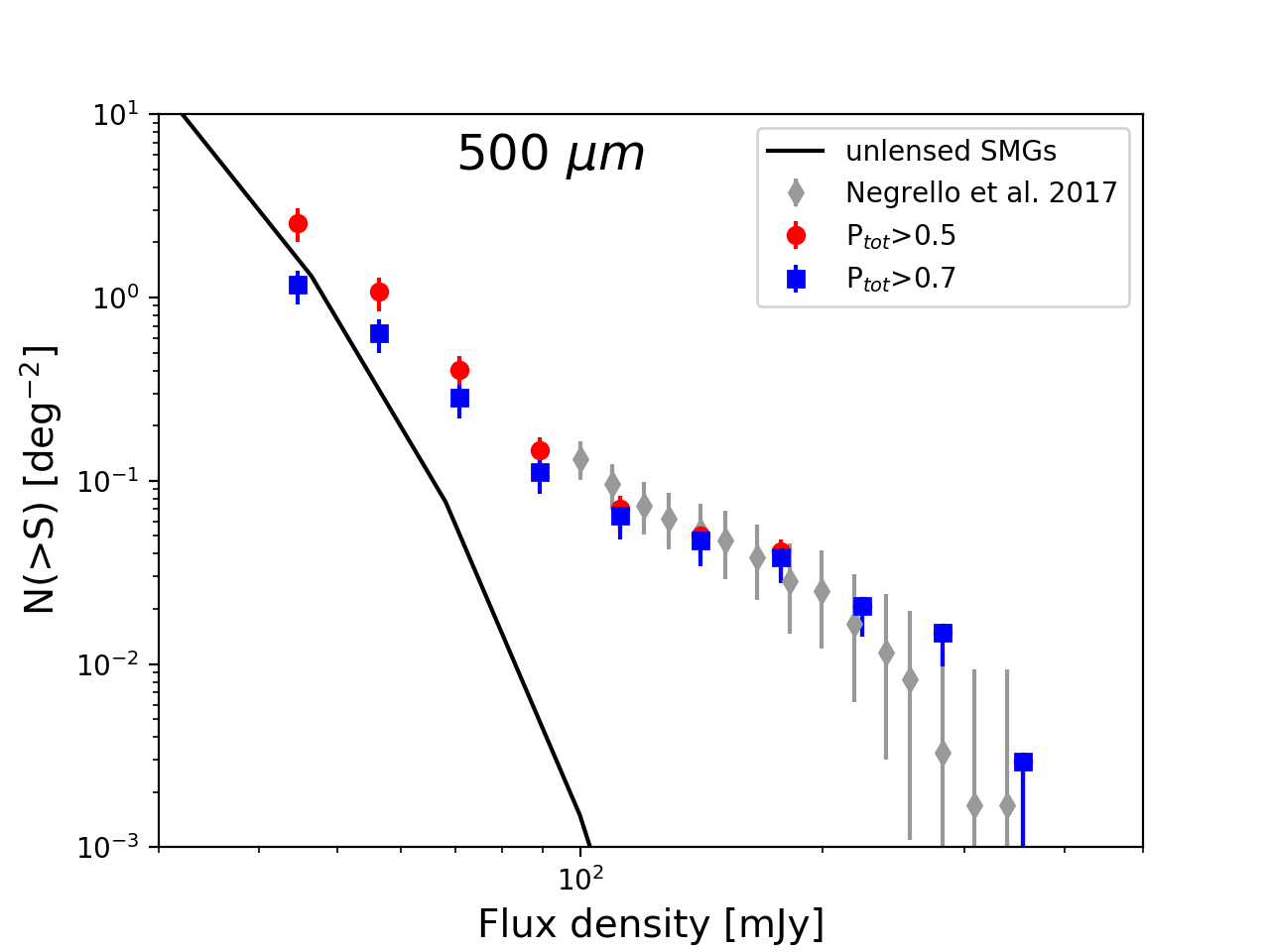}
\caption{\label{fig:snc} Integrated source number counts at $500~\mu m$ of the SHALOS candidates with P$_{tot}>0.5~\&~0.7$ (\textit{red circles} and \textit{blue squares}, respectively). They are compared with the integrated number counts of candidate lensed galaxies derived by \citet{Neg17} from all the H-ATLAS fields (\textit{grey diamonds}). The error bars correspond to the 95\% confidence interval. It is also shown the model source number counts for the unlensed SMGs \citep[\textit{black line},][]{Lap11,Cai13}.}
\end{figure}

The integral source number counts at $500\mu m$ of the SHALOS candidates, combining the results for all the four H-ATLAS fields, are shown in Fig. \ref{fig:snc}. We apply two different $P_{tot}$ cuts to check the number counts dependence on the associated probability. Above 100 mJy, we can compare the SHALOS source number counts with the one derived with the most simple but robust identification methodology by \citet{Neg17} (\textit{grey diamonds}) using the same H-ATLAS catalogues. \citet{Nay16} obtained almost identical source number counts with the same methodology but for the HeLMS+HerS survey (not shown in the Figure). Taking into account that the flux densities of the latest H-ATLAS catalogues have been updated with respect to the ones used in \citet{Neg17}, there is good agreement between both set of lensed candidates source number counts.

However, the SHALOS methodology allow us to extend the measurement of the source number counts down to 50 mJy. It is at these fainter flux densities that the effect of the different $P_{tot}$ cuts is more relevant: a lower probability cuts tend to select more lensed candidates but mainly at faiter flux limits, $S_{500~\mu m}<100$ mJy. Although we are reaching flux densities that start to be dominated by the unlensed SMGs, the SHALOS methodology seems to be effective to discriminate between the lensed/unlensed nature of the considered SMGs, at least for the $P_{tot}>0.7$ cut, as also indicated by the results of Sec. \ref{sec:acorr}.

We can conclude that the SHALOS candidates source number counts at $500~\mu m$ above 100 mJy are in good agreement with previous estimations (where many of the candidates were confirmed by follow-up observations) and, therefore, both methodologies are equivalent at such flux densities. The advantage of the SHALOS approach is that it is able to extend the identification of reliable SLG candidates down to lower flux densities, $\sim 50$ mJy.

\section{Conclusions}
\label{sec:conclusions}
We propose a new methodology for the identification of objects with particular properties by cross-matching different catalogues based on the similarity of probability distribution (the Bhattacharyya Coefficient) associated to different observables. This new approach is more simple, objective and flexible than other traditional approaches to the problem, as the LR or the Bayes factor.

As a practical application, in this work we have focused on the identification of SMGs observed by \textit{Herschel} whose flux density were strongly amplified by the gravitational lensing effect produced by SDSS galaxies at $z<0.8$, acting as the lenses. In particular, we derived the total estimated probability, $P_{tot}$, of being lensed based on four observables: the angular separation, the bolometric luminosity percentile compared with SMGs at similar redshift, the redshift difference and the ratio between the optical and the sub-mm emissions. The results indicate, as also confirmed by a PCA analysis, that the first two are the most discriminant for the identification task. The other two help to confirm that the cross-matched pairs are not the same galaxy, but two galaxies at different redshfits.

The SHALOS method identified 1451 SLG candidates with $P_{tot}>0.5$, that correspond to 0.61\% of the H-ATLAS sources. This number decrease to 447 (or 0.19\%) with a more conservative $P_{tot}>0.7$, that it is still $\sim5$ times the number of SLGs found by \citet{Neg17}. When comparing both SLGs lists, SHALOS method was able to identify 32 of the 50 SLGs with flux density at $500\mu m$ greater than $\sim90$ mJy (a lower limit to take into account small flux density variation between the different versions of the H-ATLAS catalogues). The remaining 18 SLG candidates were excluded by SHALOS because of the lack of an optical redshift estimation or for being flagged as star. On the contrary, the SHALOS method found 12 SLG candidates with $P_{tot}>0.5$ not in \citet{Neg17}: 8 have flux density at $500~\mu m$ smaller than $\sim100$ mJy, 3 are identified blazars \citep[see Table 1 in ][]{Neg17} and the last one is a local extended galaxy (NGC5705).

Finally, we have studied some characteristic statistical properties of the SHALOS SLG candidates as the estimated amplifications factors, the two-point correlation function and the source number counts. For $P_{tot}>0.7$, the tentative amplification factors were found to have mean(median) of 2.28 (1.33) for a galactic mass halo and 3.12 (1.47) for a cluster mass halo. The number of SHALOS candidates with $P_{tot}>0.7$ and $\mu>2$ are 24.3\% and 31.5\% for the galactic and cluster scenarios recpectively. 
Moreover, the SHALOS candidates have a non-zero correlation function that is stronger than the one measured for the background SMG sample in \citet{GN17}. It is in agreement with the correlation function expected for the foreground lenses \citep[massive elliptical galaxies or even group of galaxies as anticipated by \citet{GN14} and confirmed by][]{GN17}. The SHALOS candidates source number counts at $500~\mu m$ above 100mJy are in good agreement with previous results confirming that both methodologies are equivalent. However, the SHALOS one allows us to reach much lower flux densities, $\sim50$ mJy. At such faint flux densities, the total source number counts start to be dominated by unlensed SMGs, but the derived source number counts seems to indicate the effectiveness of the SHALOS methodology even in distinguishing between lensed/unlensed SMGs

\begin{acknowledgements}
JGN, LB, FA, LT and SLSG acknowledge financial support from the I+D 2015 project AYA2015-65887-P (MINECO/FEDER). JGN acknowledges financial from the Spanish MINECO for a "Ramon y Cajal" fellowship (RYC-2013-13256).
DH, FA and LT acknowledge financial support from the I+D 2015 project AYA2015-64508-P (MINECO/FEDER).
DH also acknowledges partial financial support from the RADIOFOREGROUNDS project, funded by the European Comission’s H2020 Research Infrastructures under the Grant Agreement 687312. 
JDCJ acknowledge financial support from the I+D 2017 project AYA2017-89121-P and support from the European Union's Horizon 2020 research and innovation programme under the H2020-INFRAIA-2018-2020 grant agreement No 210489629.
\\
The \textit{Herschel}-ATLAS is a project with \textit{Herschel}, which is an ESA space observatory with science instruments provided by European-led Principal Investigator consortia and with im- portant participation from NASA. The H-ATLAS website is http://www.h-atlas.org/
\\
This research has made use of \texttt{TopCat} \citep{topcat}, and the python packages \texttt{ipython} \citep{ipython}, \texttt{matplotlib} \citep{matplotlib}, \texttt{Scipy} \citep{scipy}, and \texttt{Astropy}, a community-developed core Python package for Astronomy \citep{astropy}.
\end{acknowledgements}


\bibliography{./shalos}{}
\bibliographystyle{aa.bst}

\end{document}